\documentclass{elsart}
\usepackage{graphics}

\begin{document}

\def\rhobar{\bar\rho}           
\def\erfc{\mathop{\rm erfc}}    
\def\etal{{\it{}et~al.}}        

\def\psfigure#1#2{\resizebox{#2}{!}{\includegraphics{#1}}}

\textfloatsep=1cm               
\floatsep=1.5cm                 

\date{26 November 1996}
\journal{Physica D}

\begin{frontmatter}
\title{Coherent noise, scale invariance and intermittency\\
in large systems}
\author[Nordita]{Kim Sneppen}
and
\author[SFI]{M. E. J. Newman}
\address[Nordita]{Nordita, Blegdamsvej 17, DK-2100 Copenhagen \O.
Denmark.}
\address[SFI]{Santa Fe Institute, 1399 Hyde Park Road, Santa Fe, NM 87501.
U.S.A.}
\begin{abstract}
  We introduce a new class of models in which a large number of ``agents''
  organize under the influence of an externally imposed coherent noise.
  The model shows reorganization events whose size distribution closely
  follows a power-law over many decades, even in the case where the agents
  do not interact with each other.  In addition the system displays
  ``aftershock'' events in which large disturbances are followed by a
  string of others at times which are distributed according to a $t^{-1}$
  law.  We also find that the lifetimes of the agents in the system possess
  a power-law distribution.  We explain all of these results using an
  approximate analytic treatment of the dynamics and discuss a number of
  variations on the basic model relevant to the study of particular
  physical systems.
\end{abstract}
\end{frontmatter}

\section{Introduction}
There has in the past few years been considerable interest across a broad
section of the scientific community in extended dynamical systems which
show intermittent events or ``avalanches'' of activity whose size follows a
power-law frequency distribution.  Examples of such systems include
earthquakes~\cite{GR56,KA75,PSS92,SKKV96}, solar
flares~\cite{Dennis85}, avalanches~\cite{FCMFJM96} and
laboratory experiments on stick slip dynamics~\cite{FF91}.  Behaviour
of this type presents something of a puzzle; the central limit theorem
tells us that scale-free behaviour should not exist in a system of
independent agents unless the behaviour of the individual agents is itself
scale-free, which in general it is not.  The introduction of interactions
between the agents does not necessarily alleviate the problem either.  It is
well known, for instance, that scale-free fluctuation distributions occur
in interacting systems close to continuous phase transitions.  But such
systems require careful tuning of their parameters to bring them to
criticality, whereas experiments such as those mentioned above appear to be
very robust against parameter changes.

One resolution of this difficulty was proposed by Bak, Tang and
Wiesenfeld~\cite{BTW87} who in 1987 introduced the idea of the
self-organized critical (SOC) system.  In this paper the authors proposed
that, under the action of a slow local driving force, certain systems
possessing only short-range interactions can organize themselves into a
critical state, without the need for any parameter tuning.  Once a system
organizes itself in this way, all the familiar arguments concerning
critical phenomena can be brought to bear, including the crucial concept of
universality and the robustness which it implies.  The universal properties
of critical systems, including such measurable quantities as the exponents
of power-law distributions, should be robust against changes in the
microscopic dynamics of the system; if the differences between the true
physical system---tectonic fault, sandpile or solar corona---and the simple
SOC model system can be shown (or assumed) to reside entirely in the
irrelevant variables, then the model provides not only a qualitative
description of the dynamics, but also a quantitative method for calculating
the universal properties.

The original model of Bak~\etal\ was a simple representation of the
dynamics of avalanches, but many other SOC models have been proposed since,
as models of a wide variety of physical phenomena.  Perhaps the simplest
such model is the ``minimal SOC model'' of Bak and
Sneppen~\cite{BS93}, originally constructed as a model of
coevolutionary avalanches in large ecosystems, although
Ito~\cite{Ito95} has suggested that it might also find use as a model
of earthquake dynamics.  A number of other SOC models of earthquakes have
also been proposed, including de Sousa's ``train'' version of the
sliding-block model of Burridge and Knopoff~\cite{Sousa92,BK67}, as
well as related models put forward by various
authors~\cite{TM88,IM90,CBO91,OFC92,CO92a,CO92b}.  Other applications
of the self-organized criticality idea include the modelling of forest
fires~\cite{DS92,CFO93} and the mathematically similar problem of
predator-prey interactions in spatially extended
ecosystems~\cite{SJ94}, the dynamics of solar flare
generation~\cite{LH91,LHMB93}, interface
depinning~\cite{Zaitsev92,Sneppen92}, and a variety of other
phenomena (see the review by Paczuski, Maslov and Bak~\cite{PMB96}).

However, we must also be cautious in interpreting the observation of
power-law distributions in experimental systems as evidence of SOC
behaviour.  There are many other possible explanations of how such a
power-law distribution might arise: random multiplicative
processes~\cite{MS82}, thermal crossing times for random
barriers~\cite{Ziel50}, fragmentation processes~\cite{SS91,HJS95}, and a
number of other common physical processes give rise to power-law
distributions.  One could justifiably argue that mechanisms such as these
are less flexible than SOC models, and therefore have less power as
explanations of physical phenomena.  Many of them for example can only
produce one particular value of the exponent of a power law, whereas both
SOC models and the experiments they attempt to model can show a variety of
exponent values.  However, SOC models have their limitations too.  Many SOC
models, including the minimal model of Bak and Sneppen, can be mapped in
the limit of infinite dimension onto a critical branching process, which
allows us to relate the exponent describing the fluctuation distribution to
the first return time of a one-dimensional random walk~\cite{BS93}.  The
result is that we expect these models to possess an event size distribution
characterized by an exponent of value $\threehalf$ for all dimensions above
some upper critical dimension, and lower exponent values for dimensions
below this.\footnote{In certain SOC models, particularly boundary-driven or
  deterministic ones, somewhat higher values of the exponent for the event
  size distribution have been obtained.  See for example Refs.~\cite{CO92b}
  and~\cite{IKP94}.} Unfortunately, a number of systems for which SOC
models have been proposed display distributions with steeper exponents than
this.  The distribution of the areas of terrestrial earthquakes for example
appears to follow a power-law with exponent $2.0\pm0.1$~\cite{GR56,KA75},
and the laboratory experiments on avalanches in rice-piles conducted by
Frette~\etal~\cite{FCMFJM96} show an exponent of $2.1\pm0.1$.

In this paper we discuss a new class of models which are not critical
models, but which nonetheless display a power-law distribution of event
sizes, robust against virtually any change in the microscopic dynamics of
the system.  The exponent of the power-law can take a range of values,
although for most plausible choices of the parameters of the model its
value lies in the range $2.0\pm0.2$.  The dynamics of our models is
fundamentally different from that of SOC models, being driven not by a
slow, local driving force, but by a coherent, externally-imposed noise
term.  Our models do not display long-range spatial correlation or critical
fluctuations and hence possess spatial configurations at equilibrium which
are qualitatively entirely different from SOC models.  It is our suggestion
that some of the systems traditionally studied using SOC models might in
fact be coherent-noise driven systems of the type described in this paper.

One simple example of a coherent-noise driven model displaying robust
power-law distributions has already been presented
elsewhere~\cite{NS96}.  In that paper we discussed the use of this
model as a model of sandpile dynamics and also of earthquakes.
Newman~\cite{Newman96} has also employed a variation of the model to
study the distribution of event sizes in biological extinction, another
example of a system which appears to have a power-law distribution too
steep to be easily explained using SOC dynamics.  In this paper we will
concentrate primarily on the analysis of our models, demonstrating the
origin and limits of the power-law distributions and their robustness.  The
outline of the paper is as follows.  In Section~\ref{model} we introduce
the simplest class of models displaying the phenomena of interest.  In
Section~\ref{numerical} we present numerical results of the simulation of a
large number of members of the class, displaying explicitly the robustness
of properties across the spectrum of parameter values, as well as the
limits of that robustness.  In Sections~\ref{analysis} and~\ref{beyond} we
give a detailed analysis of these models.  In Sections~\ref{discuss}
and~\ref{conclude} we discuss our results and give our conclusions.

\section{The model}
\label{model}

Consider a system composed of $N$ ``agents''.  In SOC models these agents
might be grains of sand in a sandpile, or species in an evolving ecosystem.
These agents are subject to ``stresses'' $\eta(t)$ at each time-step $t$,
chosen at random from some probability distribution $p_{\rm stress}(\eta)$.
Each agent $i$ possesses a ``threshold'' $x_i$ of tolerance for these
stresses, chosen from another distribution $p_{\rm thresh}(x)$.  If the
stress level at any time exceeds an agent's threshold value, then that
agent ``moves'', or ``dies out'', and its threshold $x_i$ is given a new
value, chosen from the same distribution.  Thus, over time, the stresses on
the system will tend to remove low threshold values from the system and
replace them with higher ones.  Assuming that the distribution $p_{\rm
  stress}(\eta)$ has a well-defined mean, this means that the system will
eventually stagnate, when we get to the point where all the thresholds are
well in excess of that mean and there are none left within reach of the
typical stresses.  In order to prevent this happening, we add a second
element to the dynamics, an ``aging'' or ``reloading'' process in which at
every time-step a certain small fraction $f$ of the agents, chosen at random
regardless of their thresholds, are given new values of $x_i$, chosen from
the same distribution as before.  The dynamics of the model can thus be
summarized as follows:

\begin{enumerate}
\item At each time-step a stress level $\eta$ is chosen at random from a
  distribution $p_{\rm stress}(\eta)$ and all agents with thresholds
  $x_i<\eta$ move and are given new threshold values chosen from a
  distribution $p_{\rm thresh}(x)$.  The number of agents which move is the
  size of the event taking place in this time-step.
\item A small fraction $f$ of the agents are chosen at random from the
  population and given new thresholds, also chosen from the distribution
  $p_{\rm thresh}(x)$.
\end{enumerate}

A few points to notice about this model:

At least in this simplest version of the model the agents are entirely
non-interacting.  We have investigated other variants in which the agents
are allowed to interact (see Section~\ref{discuss}), but the basic
predictions of the model are unchanged.

The dynamics is fundamentally a competition between the two processes.
The first tends to remove from the system agents with low thresholds for
stress.  This means that the distribution of thresholds tends to be
weighted towards high values of $x_i$.  This in turn means that the second
process, which selects agents without regard to their threshold values, is
more likely to pick ones with high $x_i$ and therefore will on average
tend to reduce the mean threshold value of the population.

The basic parameters of the system are the two distributions $p_{\rm
  stress}$ and $p_{\rm thresh}$, and the aging parameter $f$.  We have not
chosen precise values of these parameters here, and one feels intuitively
that the behaviour of the model should depend on what choice we make.  In
general, this is a correct intuition.  However---and this is our main
point---it turns out that some of the most striking properties of this
model and its variants are independent of our choice over a very large
range of possibilities.  In the case of the variable $f$ all that we will
require is that its value be small, but non-zero, in a way which we will
make precise in Section~\ref{analysis}.  The conditions on $p_{\rm stress}$
and $p_{\rm thresh}$ are more subtle, but, as we shall see, are relaxed
enough that virtually any distribution of stresses one might expect to
encounter in a real system is included.

In the next section we illustrate the basic properties of the model using
the results of numerical simulations.

\section{Numerical results}
\label{numerical}
\begin{figure}
\begin{center}
\psfigure{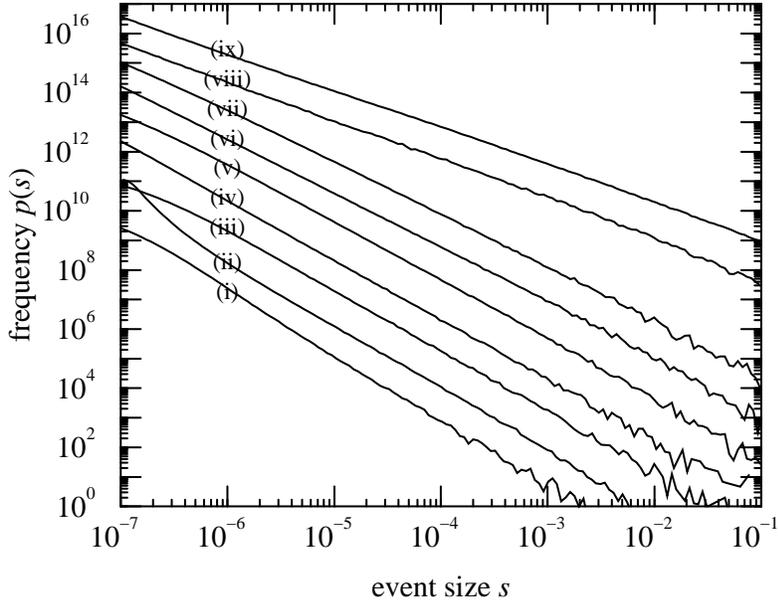}{10cm}
\end{center}
\caption{The distribution of event sizes for the model with $p_{\rm
    thresh}(x)$ uniform and a variety of choices for the imposed stress
  distribution.  The stress distributions correspond to those of Table~1 as
  follows: (i)~steep, (ii)~Gaussian~(c), (iii)~Gaussian~(b),
  (iv)~Gaussian~(a), (v)~Poissonian, (vi)~exponential, (vii)~stretched~(a),
  (viii)~stretched~(b) and (ix)~Lorentzian.  The aging parameter
  $f=10^{-6}$ in each case.  The curves have been rescaled so as not to
  overlap.
\label{avalanches1}}
\end{figure}

The most striking property of the class of models we have described is that
the distribution of the sizes of the events taking place follows a
power-law closely over many decades for most reasonable choices of $p_{\rm
  stress}$ and $p_{\rm thresh}$.  First of all, let us consider the case
where $p_{\rm thresh}(x)$ is 1 if $0\le x<1$ and zero otherwise.  In other
words, the thresholds are chosen uniformly in the interval between zero and
one.  This particular choice has the advantage that the model can be
simulated in the limit $N=\infty$ using a fast algorithm we have developed
which acts directly on the distribution $\rho(x)$ of thresholds, rather
than on the individual agents themselves.  Figure~\ref{avalanches1} shows
the results of such simulations for a variety of choices of the stress
distribution, including Poissonian noise, Gaussian noise, stretched
exponentials, and power laws.  Note that in each case, the distribution of
event sizes follows a power law over a large part of its range, deviating
only at very low event sizes.

In Table~1 we give the measured exponent $\tau$ of these power-law
distributions, and, as the table makes clear, the exponent varies depending
on the distribution of the noise.  The values are all in the vicinity of
$\tau=2$, though for very slowly-decaying noise distributions they get
closer to one.  Notice however that for systems driven by noise with the
same distribution but different values of the scale parameter $\sigma$
(e.g., the width of the Gaussian for Gaussian noise) the value of $\tau$ is
the same.  Furthermore, noise distributions with the same functional form
in their large $\eta$ tail, such as the Gaussians centered at zero and at
$\half$, also give the same values of $\tau$.

\begin{table}
\begin{center}
\begin{tabular}{ccc}
noise type & $p_{\rm stress}(\eta)$ & exponent $\tau$ \\
\hline
\hline
steep          & $\exp(-\eta^4/\sigma^4)$                     & $2.22 \pm 0.06$ \\
Gaussian (a)   & $\exp(-\eta^2/2\sigma^2)$ with $\sigma=0.05$ & $2.01 \pm 0.02$ \\
Gaussian (b)   & $\exp(-\eta^2/2\sigma^2)$ with $\sigma=0.1$  & $2.04 \pm 0.02$ \\
Gaussian (c)   & $\half + \exp(-\eta^2/2\sigma^2)$            & $2.04 \pm 0.04$ \\
Poissonian     & $\mu^{\eta/\sigma}/\Gamma(\eta/\sigma+1)$    & $1.97 \pm 0.03$ \\
exponential    & $\exp(-\eta/\sigma)$                         & $1.85 \pm 0.03$ \\
stretched (a)  & $\exp[-(\eta/\sigma)^{0.5}]$                 & $1.76 \pm 0.04$ \\
stretched (b)  & $\exp[-(\eta/\sigma)^{0.2}]$                 & $1.32 \pm 0.02$ \\
Lorentzian     & $(\sigma^2 + \eta^2)^{-1}$                   & $1.30 \pm 0.08$ \\
\hline
\hline
\end{tabular}
\end{center}
\medskip
\caption{Definition of the various noise distributions used in
  Figure~\protect\ref{avalanches1}, along with the measured value of the
  event size exponent $\tau$ for each case.}
\end{table}

\begin{figure}
\begin{center}
\psfigure{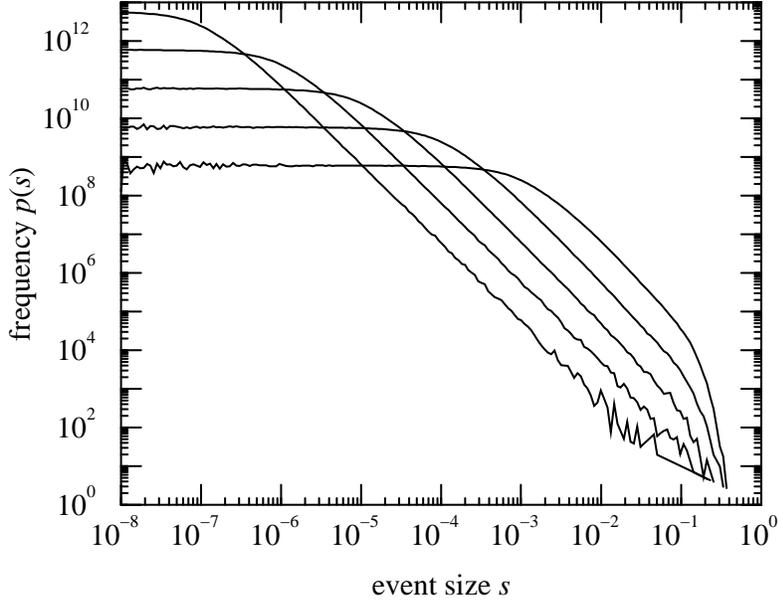}{10cm}
\end{center}
\caption{The distribution of event sizes for the model with
  a Gaussian stress distribution of mean zero and width $\sigma=0.05$,
  with values for the aging parameter $f$ varying from $10^{-6}$
  (lowest curve) to $10^{-2}$ (highest).
\label{avalanches2}}
\end{figure}

In Figure~\ref{avalanches2} we show the results of a set of simulations for
systems with the same Gaussian noise distribution, but different values
of the aging parameter $f$.  As the figure shows, varying $f$ varies the
size of the region over which power-law behaviour is seen, but again, does
not vary the exponent of the power-law.

\begin{figure}
\begin{center}
\psfigure{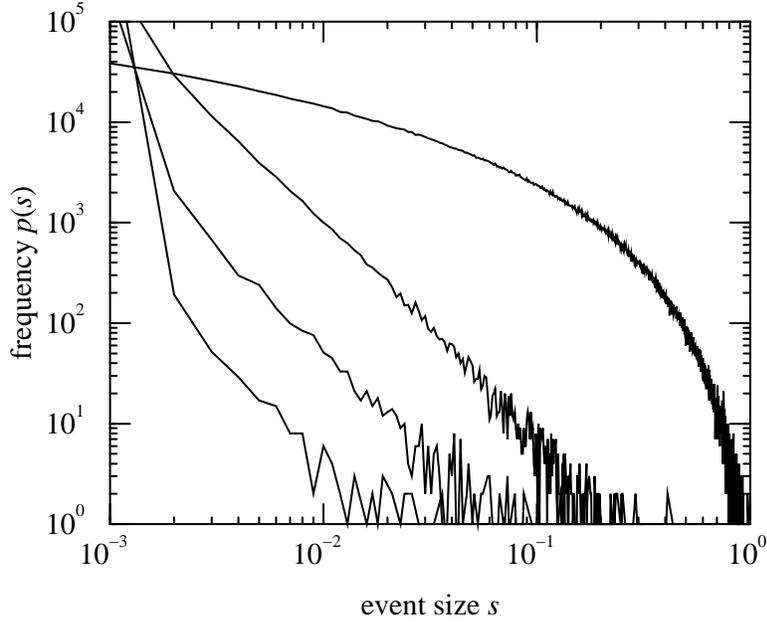}{10cm}
\end{center}
\caption{The distribution of event sizes for the model with
  stresses distributed exponentially (see Table~1) with $\sigma$ held fixed
  at $0.05$ and an aging parameter $f=10^{-6}$.  The different curves
  correspond to different choices of the distribution $p_{\rm thresh}(x)$
  from which the thresholds are selected.  In this case $p_{\rm
    thresh}(x)$ was also an exponential distribution with decay constants
  (left to right) of $0.5$, $0.2$, $0.1$ and $0.05$.  The power-law form is
  maintained until the decay constant becomes extremely short.
\label{avalanches3}}
\end{figure}

In Figure~\ref{avalanches3} we show results for simulations performed with
choices of $p_{\rm thresh}$ different from the uniform one used above.  In
this case we chose both $p_{\rm thresh}$ and $p_{\rm stress}$ to be
exponentials, and held the decay parameter $\sigma$ for $p_{\rm stress}$
fixed whilst varying that for $p_{\rm thresh}$.  As long as the fall-off in
$p_{\rm thresh}$ is slower than that for $p_{\rm stress}$, we continue to
see power-law behaviour.  But once $p_{\rm thresh}$ starts to fall off
faster, the power-law is destroyed.  (Note that these simulations were,
necessarily, performed for finite $N$, using a slower algorithm than the
previous figures, so the statistics are poorer.)

\begin{figure}
\begin{center}
\psfigure{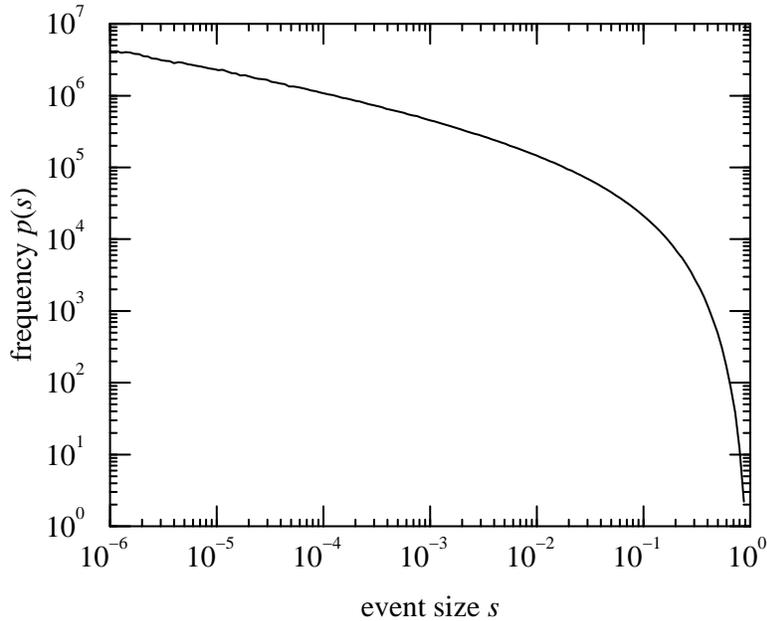}{10cm}
\end{center}
\caption{The distribution of event sizes for a simulation in which both the
  thresholds and the stress levels were chosen uniformly in the interval
  between zero and one.  Clearly this distribution does not obey a power
  law.
\label{avalanches4}}
\end{figure}

In Figure~\ref{avalanches4} we show the distribution of events for a system
in which the stresses were chosen uniformly in the interval between zero
and one, just as were the thresholds.  Clearly, this choice does not result
in a power-law distribution of event sizes.

In Section~\ref{analysis} we offer theoretical arguments in explanation of
each of these phenomena.

\begin{figure}
\begin{center}
\psfigure{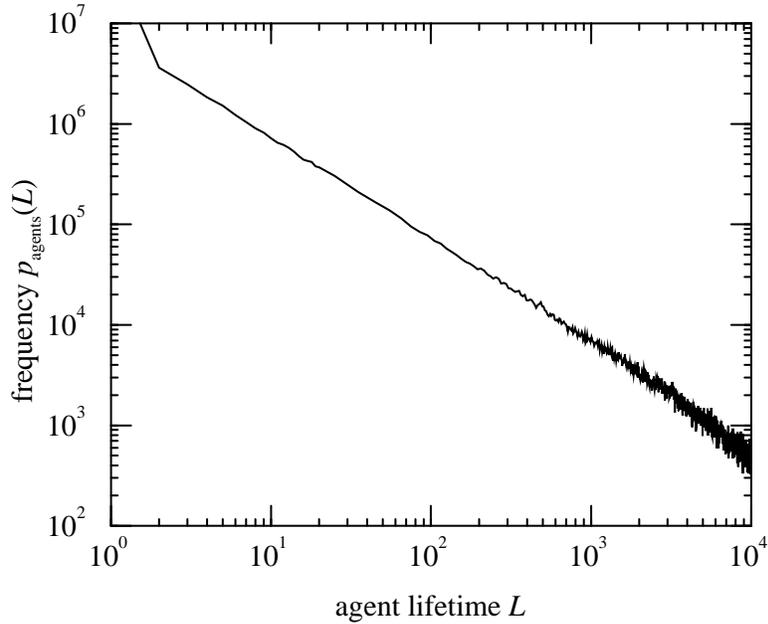}{10cm}
\end{center}
\caption{The distribution of lifetimes of agents in a simulation with
  exponential stresses.  The distribution is a power law with an exponent
  of $1.00\pm0.02$.
\label{lifetimes}}
\end{figure}

There are a number of other interesting features to be seen in the
behaviour of these models.  In Figure~\ref{lifetimes}, we show the
distribution of the ``lifetimes'' of agents in the model---the number of
time-steps between one move and the next of a particular agent.  This too
is a power-law, with an exponent close to $1$ in the case of the
exponential stress distribution employed here.  (Another measure of the
time behaviour of the system is the power spectrum of the pattern of
activity.  However, this spectrum turns out to go like $1/f^2$, a form seen
in such a wide variety of dynamical systems that it tells us very little
about the properties of the model.)

\begin{figure}
\begin{center}
\psfigure{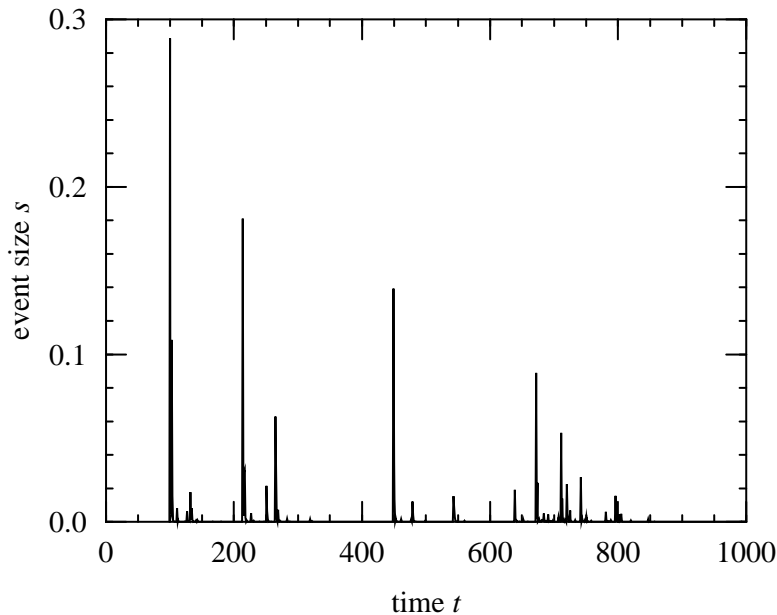}{10cm}
\end{center}
\caption{A section of the time-series of reorganization events in a
  simulation of the model.  The aftershocks following the first large
  event are clearly visible.  Notice also that some of the aftershocks
  themselves generate a smaller series of after-aftershocks.
\label{timeseries}}
\end{figure}

\begin{figure}
\begin{center}
\psfigure{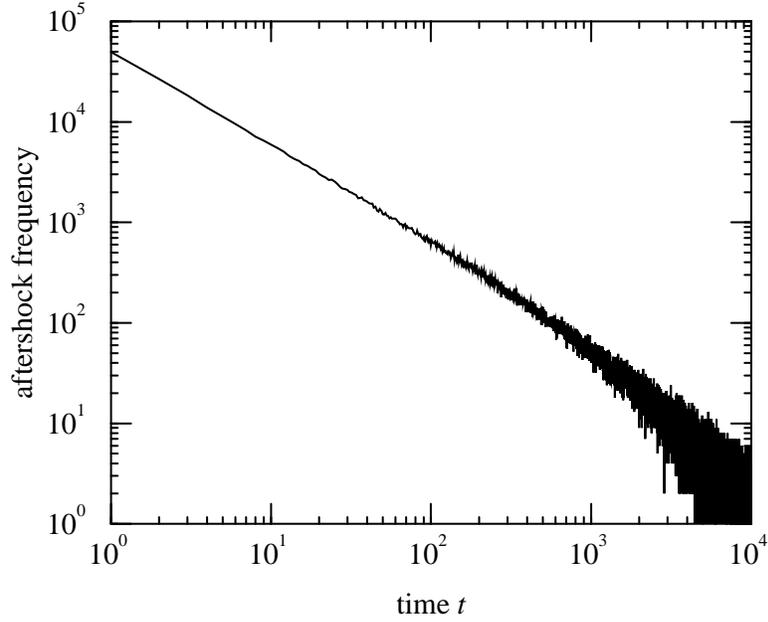}{10cm}
\end{center}
\caption{A histogram of the time distribution of aftershocks following a
  large event.  The distribution follows a power law with an exponent
  very close to one.
\label{aftershocks}}
\end{figure}

A feature of these models which may prove useful in distinguishing the
present dynamics from SOC systems is the existence of
``aftershocks''---large events following closely on the tail of others.
Figure~\ref{timeseries} shows a sample of time-series data from one of our
simulations displaying a set of aftershocks.  In Figure~\ref{aftershocks}
we have measured the time-delay between a large event and each of the
succeeding aftershocks.  A histogram of these time delays shows a $t^{-1}$
functional form.  Interestingly the exponent here seems to be independent
of the nature of the stresses driving the system.

\section{Analysis}
\label{analysis}
As we saw in the last section, models of the type discussed here generate
scale-free distributions of a number of measurable quantities, including
the sizes of the reorganization events.  These distributions are robust
against most changes we can make in the parameters of the models, and for
this reason we believe that models of this type may provide an explanation
for the characteristic fluctuation distributions and intermittency seen in
a number of extended dynamical systems.  As Figures~\ref{avalanches3}
and~\ref{avalanches4} show however, there are limits to the robustness of
the various distributions, as well as some qualitative differences between
these models and the SOC models (such as the aftershocks depicted in
Figure~\ref{timeseries}).  In this section we develop a theoretical picture
of the behaviour of our models, demonstrate the causes of the robustness
and the reasons why it breaks down in some cases, and give quantitative
explanations of the existence and distribution of aftershocks, agent
lifetimes, and other features of these systems.

First, we note that it is a straightforward matter to solve for the mean
(i.e.,~time-averaged) distribution $\rhobar(x)$ of the threshold variables
$x_i$.  In any small interval $\d x$, the rate at which agents leave
the interval because of the aging process is $f\rhobar(x)\>\d x$.  In the same
interval the probability per unit time $p_{\rm move}(x)$ of an agent being
hit by a stress event is simply the probability that the stress level
$\eta$ will be greater than $x$ in that time-step, which is
\begin{equation}
p_{\rm move}(x) = \int_x^\infty p_{\rm stress}(\eta)\>\d\eta.
\label{pmove}
\end{equation}
The mean rate at which agents in the interval get hit by the stress events
is therefore equal to $p_{\rm move}(x)\>\rhobar(x)\>\d x$.  The rate at which
the interval $\d x$ is repopulated is simply $A\> p_{\rm thresh}(x)\>\d x$,
where $A$ is an $x$-independent constant whose value is equal to the total
integrated rate at which agents are lost to both the aging and stress
processes.  Thus, we can write a master equation balancing the rates of
agent loss and repopulation which should describe the time-averaged
equilibrium state of the model:
\begin{equation}
f\rhobar(x) + p_{\rm move}(x)\>\rhobar(x) = A\>p_{\rm thresh}(x).
\label{master}
\end{equation}
Rearranging for $\rhobar(x)$, we get
\begin{equation}
\rhobar(x) = {A\>p_{\rm thresh}(x)\over f+p_{\rm move}(x)}.
\label{rhobar}
\end{equation}
This equation is exact.  For any given choice of $p_{\rm thresh}(x)$ and
$p_{\rm stress}(\eta)$, the constant $A$ can be fixed by requiring that
$\rhobar(x)$ integrate to unity over the allowed range of $x$ values.

\begin{figure}
\begin{center}
\psfigure{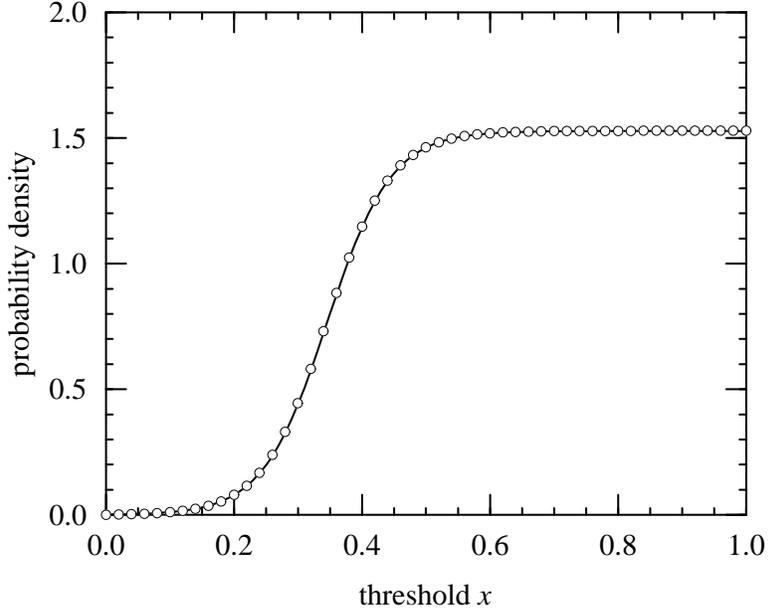}{10cm}
\end{center}
\caption{The average threshold distribution $\rhobar(x)$ for the model with
  exponentially distributed stresses with $\sigma=0.05$.  The solid line is
  the analytic expression, Equation~(\ref{rhobar}), and the points are the
  numerical data.
\label{threshold}}
\end{figure}

To give a concrete example, consider the simple case where $p_{\rm
thresh}(x)$ is uniformly distributed over the interval between zero and one
and $p_{\rm stress}(\eta)$ is the normalized exponential distribution
\begin{equation}
p_{\rm stress}(\eta) = \frac{1}{\sigma} \exp(-\eta/\sigma).
\label{exponential}
\end{equation}
(The stress $\eta$ is restricted to non-negative values.)  In this case,
$p_{\rm move}(x) = \exp(-x/\sigma)$ and so
\begin{equation}
\rhobar(x) = {A\over f+\exp(-x/\sigma)}.
\end{equation}
The constant $A$ is
\begin{equation}
A = {f\over\sigma} \biggl[ \log{f\exp(1/\sigma)+1\over f+1} \biggr]^{-1}.
\label{normalization}
\end{equation}
In Figure~\ref{threshold} we have plotted this expression for $\rhobar(x)$
along with data for the same quantity from our numerical studies.  The
agreement between the two is excellent.  Notice some crucial features of
the curve.  First, the probability $p_{\rm move}(x)$ that an agent with
threshold $x$ will lie below the stress level on any particular time-step
decreases monotonically with $x$, regardless of the form of the stress
distribution.  Thus, as long as $f$ is finite there will always be some
value of $x$, call it $x_c$, above which Equation~(\ref{rhobar}) is
dominated by the first term in the denominator so that $\rhobar(x)$ is
simply proportional to $p_{\rm thresh}(x)$.  Below $x_c$ there is a regime
in which the second term dominates, making $\rhobar(x)$ proportional to
$p_{\rm thresh}(x)/p_{\rm move}(x)$:
\begin{equation}
\rhobar(x) \approx {A\> p_{\rm thresh}(x)\over p_{\rm move}(x)}.
\label{approxrhobar}
\end{equation}
In the exponential example considered above this becomes
\begin{equation}
\rhobar(x) \approx A\>\exp(x/\sigma).
\end{equation}
As we will see, it is this part of the curve which is responsible for the
power-law distribution of event sizes in the model.  The crossover point
$x_c$ between the two regimes is defined by
\begin{equation}
p_{\rm move}(x_c) = f,
\label{crossover}
\end{equation}
which means that the value of $f$ controls the range of event sizes over
which we have power-law behaviour, the range increasing as $f$ gets
smaller.

The size of the event which corresponds to a stress of magnitude $\eta$ is
given by
\begin{equation}
s(\eta) = \int_{-\infty}^\eta \rho(x)\>\d x.
\label{sofeta}
\end{equation}
In general the threshold distribution at any particular time $t$ will
differ from the time-averaged distribution $\rhobar(x)$.  In this section
however, we will make the approximation that the two are equal, which
allows us to solve for the event size distribution.  We will refer to this
as the ``time-averaged approximation'' (TAA), and in practice it turns out
to be a good guide to the average behaviour of the system, though there are
some limitations to what it can tell us.  (In particular, it can tell us
nothing about time correlations in the system, and therefore cannot explain
the aftershocks of Figure~\ref{timeseries}, for example.)

For thresholds distributed according to Equation~(\ref{rhobar}), the
probability density of events of size $s$ is given by
\begin{equation}
p_{\rm event}(s) = p_{\rm stress}(\eta) {\d\eta\over\d s} =
                  {p_{\rm stress}(\eta(s))\over \rhobar(\eta(s))},
\label{mft}
\end{equation}
where we have calculated $\d\eta\over\d s$ by substituting $\rhobar(x)$ in
place of $\rho(x)$ in Equation~(\ref{sofeta}) and differentiating, and the
function $\eta(s)$, which is the stress required to produce an event of
size $s$, is given by the functional inverse of Equation~(\ref{sofeta}).
This function $\eta(s)$ is a monotonic increasing function of $s$, and this
immediately leads us to our first result about the model: all events of
size greater than a given size $s$ are produced by stresses greater than
$\eta(s)$.  Thus, the high-$s$ tail of the event distribution depends only
on the functional form of the high-$\eta$ tail of the stress distribution.
This offers an explanation (at least within the TAA) of the behaviour
observed in the figures of Section~\ref{numerical}, in which stress
distributions with the same high-$\eta$ tail gave rise to event
distributions obeying the same power-law.

We are also now in a position to answer the question of how the power-law
distribution of events arises.  As we mentioned above, the power-law
depends on the low-$x$ part of the threshold distribution, the part below
$x_c$.  In this regime Equation~(\ref{approxrhobar}) is valid and, making
use of Equations~(\ref{pmove}) and~(\ref{mft}), we write
\begin{equation}
p_{\rm event}(s) = A^{-1} {p_{\rm stress}(\eta(s)) \int_{\eta(s)}^\infty p_{\rm
    stress}(x)\>\d x \over p_{\rm thresh}(\eta(s))},
\label{aval}
\end{equation}
Making the same approximations in Equation~(\ref{sofeta}), we get
\begin{equation}
s(\eta) = \int_{-\infty}^\eta {A\>p_{\rm thresh}(x)\over\int_x^\infty p_{\rm
    stress}(\eta')\>\d\eta'} \>\d x.
\label{size}
\end{equation}
Between them, these two equations define the event size distribution.  As it
turns out the form of $p_{\rm thresh}(x)$ is virtually irrelevant to the
form of $p_{\rm event}(s)$, our only requirement being that it should vary
relatively little between the minimum value of $x$ and the value $x_c$.
Variation by a factor of ten or so would be acceptable.  To see this we
need only consider the distributions of event sizes shown in the last
section: typically $p_{\rm event}(s)$ varies over many orders of magnitude,
as many as twenty in some cases.  Clearly then, one order of magnitude
variation in the denominator of Equation~(\ref{aval}) is not going to make
much difference.  It is certainly possible to arrange for $p_{\rm
  thresh}(x)$ to vary fast enough to destroy the power-law behaviour---just
as we did in Figure~\ref{avalanches3}---but for most reasonable choices the
functional form is unimportant.

So what does the event distribution depend on?  The crucial condition
which must be fulfilled if we are to get a power-law distribution of
event sizes turns out to be that the value of $p_{\rm stress}(\eta)$ in
the tail should fall off fast enough that the integral of $p_{\rm
  stress}(\eta)$ from $\eta$ to $\infty$ should be dominated by the value
of the function near $\eta$.  In its most general form, this condition can
be stated as:
\begin{equation}
\int_\eta^\infty p_{\rm stress}(x)\>\d x \sim p_{\rm stress}^\alpha(\eta).
\label{condition}
\end{equation}
Substituting this condition into Equations~(\ref{aval}) and~(\ref{size}),
and neglecting variation in $p_{\rm thresh}(x)$, we find that
\begin{equation}
p_{\rm event}(s) \sim p_{\rm stress}^{\alpha+1}(\eta(s)),
\end{equation}
and
\begin{eqnarray}
s(\eta) &\sim& \int_{-\infty}^\eta p_{\rm stress}(x)^{-\alpha}\>\d x
        = \int_0^{p_{\rm stress}(\eta)} p_{\rm stress}^{-\alpha}
            {\d x\over\d p_{\rm stress}} \>\d p_{\rm stress}\nonumber\\
        &=& {1\over p_{\rm stress}(\eta)}, 
\end{eqnarray}
where we have employed Equation~(\ref{condition}) to evaluate the
derivative.  Now we can combine these two equations to give:
\begin{equation}
p_{\rm event}(s) \sim s^{-\tau},
\end{equation}
where
\begin{equation}
\tau = \alpha + 1.
\end{equation}
As we will see, the value of $\alpha$ for most of the distributions $p_{\rm
stress}(x)$ which we have considered is close to one, with the result that
$\tau$ is usually close to two.

To make these developments a little more specific, let us look at some
examples.  Two cases in which Equation~(\ref{condition}) is exactly true
are the cases of stress distributions $p_{\rm stress}(\eta)$ with either an
exponential or a power-law tail:
\begin{eqnarray}
\int_\eta^\infty \e^{-x/\sigma} \>\d x \propto
\e^{-\eta/\sigma}&,&\qquad\mbox{i.e.,~$\alpha=1$}\\
\int_\eta^\infty \Bigl({x\over\sigma}\Bigr)^{-\gamma} \>\d x \propto
       \Bigl({\eta\over\sigma}\Bigr)^{-\gamma+1}&,&
       \qquad\mbox{i.e.,~$\alpha=1-\frac{1}{\gamma}$.}
\end{eqnarray}
Hence, we can expect the distributions of events produced by, for example,
the Poissonian and Lorentzian stress distributions to be essentially
perfect power-laws.  In almost all the other cases considered in the last
section, the condition~(\ref{condition}) is true to a good approximation,
which explains why we still see power-law behaviour over many decades.  For
example, in the case of stress distributions with a Gaussian tail we have:
\begin{equation}
\int_\eta^\infty \exp\Bigl(-{x^2\over\sigma^2}\Bigr) \>\d x \propto
       \erfc(\eta/\sigma).
\end{equation}
However, the error function $\erfc(x)$ has a tail which goes like
$\exp(-x^2)/x$.  The $x^{-1}$ decays much slower than the Gaussian, and so
to a good approximation Equation~(\ref{condition}) still holds over any
region in which $\eta$ is comparable to or greater than $\sigma$.  Thus in
the case of Gaussian stresses we also find a power-law distribution of
event sizes.  The theory outlined above predicts $\tau=2$ in both the
exponential and Gaussian cases, and $\tau=\threehalf$ in the Lorentzian
case.  These figures are in moderately good agreement with measured figures
of $\tau=2.0$ for the Gaussian, $\tau=1.9$ for the exponential, and
$\tau=1.3$ for the Lorentzian.  For most of the other cases examined in
Section~\ref{numerical} we can also demonstrate, at least approximately,
the validity of Equation~(\ref{condition}) and deduce a value for the
exponent $\tau$ which is in approximate agreement with the numerical
results.  The extent to which the predicted and observed exponents do not
agree is a measure of the validity of the TAA; the difference between the
two is presumably due to fluctuation of $\rho(x)$ around the mean
distribution $\rhobar(x)$.

Stress distributions which do not satisfy Equation~(\ref{condition}), such
as the flat one used in Figure~\ref{avalanches4}, or indeed any
distribution which does not have a tail, will not produce a power-law
distribution of event sizes.  However, we contend that virtually all
naturally occurring stress distributions which a system is likely to
encounter are covered by the many cases we have shown in
Section~\ref{numerical}, and hence that, for systems driven by coherent
stress in this fashion, power-law event size distributions with exponents
in the vicinity of two should be a ubiquitous phenomenon.

Now we turn our attention to the distribution of lifetimes of the agents in
the model.  Using the same approximations which we employed above to study
the event size distribution we can show that we should also expect the
lifetime distribution to follow a power law.  First, note that the typical
lifetime $L$ of an agent possessing a threshold value $x$ is simply related
to the probability $p_{\rm move}(x)$ (Equation~(\ref{pmove})) of it being
hit by the stress on any particular time-step:
\begin{equation}
L = {-1\over\log[1-p_{\rm move}(x)]} \approx {1\over p_{\rm move}(x)},
\label{lifetime}
\end{equation}
the latter equality holding when $L$ is significantly greater than one,
which is the case we are interested in.  We can then use this equation to
relate the probability distribution of lifetimes over the population
$p_{\rm pop}(L)$ to the distribution of thresholds $x$, which we
approximate by its value within the TAA:
\begin{equation}
p_{\rm pop}(L) = \rhobar(x) {\d x\over\d L}.
\end{equation}
Note however, that the contribution of the longer-lived agents to this
distribution will be greater than that of the shorter-lived ones, in
proportion to their lifetimes.  It is more common to calculate the
distribution of lifetimes as an average simply over all agents, a
distribution which would be given by:
\begin{equation}
p_{\rm agents}(L) \propto \rhobar(x) {1\over L} {\d x\over\d L}.
\end{equation}
The derivative can be evaluated using Equations~(\ref{pmove})
and~(\ref{lifetime}) thus:
\begin{equation}
{\d L\over\d x} = {-1\over p_{\rm move}^2(x)} {\d p_{\rm move}(x)\over\d x}
                = {p_{\rm stress}(x)\over p_{\rm move}^2(x)}.
\end{equation}
The threshold distribution $\rhobar(x)$ we get from
Equation~(\ref{approxrhobar}), assuming as we did earlier that $p_{\rm
thresh}(x)$ changes little over the region of interest.  Then
\begin{equation}
p_{\rm agents}(L) \propto {1\over L} \> {p_{\rm move}(x)\over p_{\rm
stress}(x)} \sim L^{-2+\frac{1}{\alpha}},
\label{lifedist}
\end{equation}
where we have made use of Equation~(\ref{condition}) again.  For stress
distributions such as the exponential, Poissonian and Gaussian
distributions, for which $\alpha=1$, this gives a lifetime distribution
which goes like $L^{-1}$, a figure which is confirmed well by the numerical
data.  The exponent in Figure~\ref{lifetimes}, for example, is $1.00\pm0.02$.

\section{Results beyond the time-averaged approximation}
\label{beyond}
One result which is clearly outside the realm of the time-averaged
approximation, but which nevertheless admits of a very simple explanation,
is the existence of the aftershock events seen in Figure~\ref{timeseries},
and the power-law distribution of the times at which they occur, as shown
in Figure~\ref{aftershocks}.  The basic explanation of the existence of the
aftershocks is this.  When a large stress event occurs, a significant
fraction of the agents in the system move, and acquire new threshold values
distributed according to $p_{\rm thresh}(x)$.  As we pointed out in
Section~\ref{model}, this typically lowers the mean threshold value,
producing a distribution $\rho(x)$ with more weight at low $x$ than the
average distribution $\rhobar(x)$.  The result is that when, some short
time later, another stress of only moderate size hits the system, it will
affect a larger number of agents than one would normally expect, producing
an aftershock event.

As Figure~\ref{aftershocks} shows, the times $t$ after the initial large
event at which aftershocks occur, are distributed according to a power law.
The exponent of this distribution is one in all cases we have investigated,
a finding which we explain as follows.  Consider what happens when a large
event takes place.  Following such an event, a large fraction of the agents
in the system are assigned new threshold values drawn at random from the
distribution $p_{\rm thresh}(x)$.  Shortly thereafter, the first aftershock
event occurs, when some moderate-sized stress $\eta_1$ hits the system and
wipes out all agents with $x<\eta_1$.  In order to get a second aftershock
we now need a larger stress $\eta_2$, to reach agents who were not hit
during the first aftershock.  If it took a time $t_1$ to generate a stress
of size $\eta_1$, then it will, on average, take as long again to generate
another stress of greater or equal magnitude, or an aggregate time of $t_2
= 2t_1$.  Repeating this argument for the third aftershock we can see that
it will occur an average of $t_3 = 2t_2$ after the initial large event, and
so forth.  In general the $n^{\rm th}$ aftershock will occur a time
\begin{equation}
t = t_1 2^n
\label{aftertime}
\end{equation}
after the initial event.  The average number of aftershocks $\d n$ occurring
in an interval $\d t$ of time after the initial large event (which is what
Figure~\ref{aftershocks} shows) is then
\begin{equation}
\d n = {\d t\over t_1 n 2^{n-1}} = {2\d t\over n t}.
\end{equation}
From Equation~(\ref{aftertime}) we can see that the $n$ in the numerator
varies as $\log t$.  Thus, apart from logarithmic corrections, the
distribution of aftershock times goes as $t^{-1}$.  This argument makes no
assumptions at all about the form of $p_{\rm stress}(\eta)$ or $p_{\rm
thresh}(x)$, and so can be expected to hold regardless of how we choose
these functions, which is indeed what we see in our simulations.

\begin{figure}
\begin{center}
\psfigure{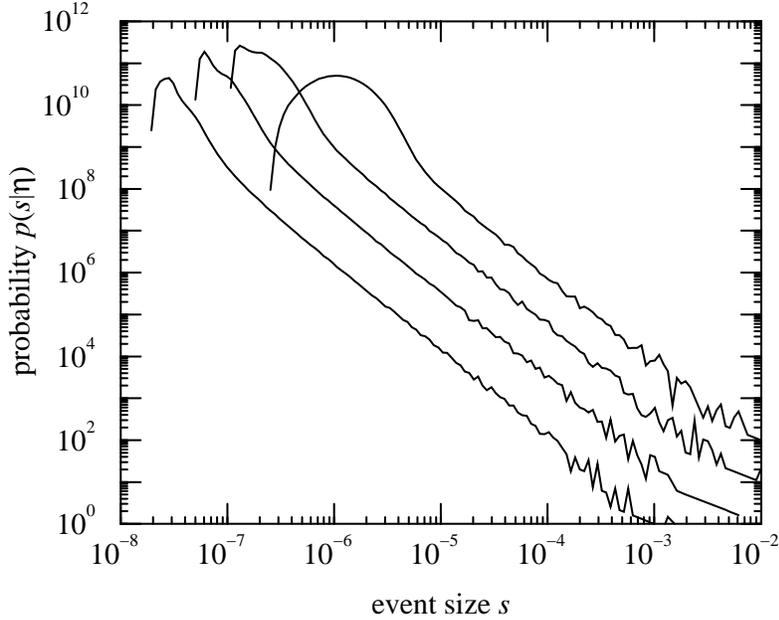}{10cm}
\end{center}
\caption{A plot of the probability $p(s|\eta)$ of getting an event of size
  $s$ given a stress of size $\eta$.  The stresses were Gaussianly
  distributed with mean zero and width $\sigma=0.1$, the aging parameter
  $f=10^{-6}$, and the values of $\eta$ are (from left to right) $0.02$,
  $0.05$, $0.1$ and $0.2$.
\label{psgiven}}
\end{figure}

We can also ask about the size of the aftershock events, and it turns out
that there is a very simple argument giving the distribution of this
quantity as well.  Consider a large event which reorganizes a significant
fraction $s_0$ of the agents in the system.  The disturbed agents are
redistributed according to the function $p_{\rm thresh}(x)$, which, as we
have assumed, is slowly varying over the region of $x$ of interest.  If our
large event is followed some time later by a stress event of moderate size
$\eta_1$, the size $s_1$ of the corresponding aftershock is then
proportional to $s_0$.  Given that our TAA predicts that $s_0$ will be
distributed according to a power law, this tells us that the aftershocks
must be distributed according to the same power law, even though the TAA
itself can tell us nothing about the aftershocks.  We can extend this
argument to tell us something about the distribution $p(s|\eta)$ of event
sizes $s$ produced by stresses of a certain strength $\eta$: the large-$s$
tail of $p(s|\eta)$ is presumably dominated by aftershock events since
these are, on average, larger than events for the same size $\eta$ which do
not occur immediately after another large event.  But the distribution of
aftershocks has, as we have said, a power-law form with the same exponent
as the overall event size distribution, and hence we can conclude that
$p(s|\eta)$ should also fall off like a power law with the same exponent.
In Figure~\ref{psgiven} we have plotted $p(s|\eta)$ for a variety of values
of $\eta$ on logarithmic scales which clearly shows this power-law
functional form.  Within the TAA, we assume that only one size of event is
produced by a given value of $\eta$ (see Equation~(\ref{sofeta})) and hence
that $p(s|\eta)$ is equal to a $\delta$-function $\delta(s-s(\eta))$, but
this is only an approximation.  In reality, as Figure~\ref{psgiven} shows,
a stress of size $\eta$ can produce a variety of different event sizes,
depending on the instantaneous distribution of thresholds, but, to the
extent that this distribution is narrow and approximates well to a
$\delta$-function, the TAA should be a reliable approximation.

\section{Discussion}
\label{discuss}
In the previous two sections of this paper, we have seen that, subject to
certain conditions, the class of models described in Section~\ref{model}
should show robust scale-free distributions of event sizes, agent
lifetimes, and aftershocks.  The conditions are:

(i) The value of $f$ should be small.  This ensures that the first term in
the denominator of Equation~(\ref{rhobar}) can be neglected.

(ii) The distribution $p_{\rm thresh}(x)$ from which thresholds are
selected should not vary too much over the range of $x$ values up to the
crossover point $x_c$ defined by Equation~(\ref{crossover}).  In particular
its dynamic range should be much smaller than that of the event
distribution $p_{\rm event}(s)$ over the power-law portion of its domain.
In the cases examined in Section~\ref{numerical} this condition is easily
fulfilled, the event distribution having a dynamic range running to ten
orders of magnitude or more.  Furthermore, this dynamic range can be
increased by the simple expedient of decreasing $f$, so there will always
be a regime in which this condition is satisfied.

(iii) Most importantly, the condition embodied in
Equation~(\ref{condition}) should be obeyed.  It is this condition which
governs the stress distributions $p_{\rm stress}(\eta)$ which produce
power-law event size distributions.  Equation~(\ref{condition}) is obeyed,
at least approximately, by any distribution whose tail falls off as an
exponential, Gaussian, stretched exponential or power law, and in all these
cases our numerical results do indeed confirm the existence of a power-law
event distribution.  In our opinion this list covers essentially all stress
distributions one is likely to find in naturally occurring systems, making
power laws the generic response of large systems to coherent noise.

We would like to stress that the root cause of the power-law distributions
seen in these models is not related to critical phenomena.  These models
are not critical in any usual sense of the word.  In the simple form we
have discussed so far, the agents comprising these systems do not even
interact with one another---the dynamics of each would be unaffected if we
were to take all the others away---and hence there can be no correlated
fluctuations of the type which are normally taken as definitive of critical
systems.  The organized nature of the fluctuations in our models arises
instead because each agent feels the same stress at the same time.

One feature of our models which is in particularly striking contrast to SOC
systems is the aftershocks discussed at the end of the last section.  These
aftershocks arise as a result of the memory the system has of events in its
recent past.  SOC models also have a memory of their history, but the
states of the system before and after an event are statistically identical,
with the result that the probability of a moderate-sized event occurring is
no higher immediately after a large event than at any other time.
This suggests that the existence of aftershocks could provide a useful
measure for distinguishing between coherent-noise driven systems of the
type considered here, and SOC dynamics.  (The aftershocks should not be
confused with the effect seen in some SOC models, in which the chance of a
given site being hit during a particular avalanche is enhanced if that site
has been hit previously in the same avalanche.  This effect does not result
in a string of small avalanches in the wake of a large one, as our
aftershocks do.)


As we mentioned at the beginning of Section~\ref{model}, many
variations are possible on the basic theme of a coherent-noise driven
model.  Before closing, let us examine briefly a couple of the more
interesting options we have investigated.  One twist on the basic idea
which may be of relevance to the modelling of earthquakes is a spatial
version which goes like this.  Our agents are now put on a lattice and
during every time-step of the model we move not only those agents with
threshold values lying below the instantaneous stress level, but also
their immediate neighbours on the lattice.  This mechanism is
interesting because it introduces correlations in the spatial
distribution of successive events: the neighbouring agents which get
moved during one event will on average have lower thresholds than
those that remained where they were, with the result that the agents
which are struck down in succeeding time-steps will tend to be
clustered around those of previous time-steps.  However, although this
might potentially provide a way of modelling the spatial distribution
of earthquakes or reorganization events in other systems, it does not,
as far as we are able to make out, affect the distribution of event
sizes, or the other distributions we have measured.  Thus we believe
the scaling properties of our model to be robust under changes of this
sort to the dynamics of the system.\footnote{Another slight variation
  on this idea, which might have applications in the modelling of the
  surface dynamics of avalanches (such as the one-dimensional
  rice-pile experiments of Frette~\etal~\cite{FCMFJM96}), is one in
  which the agents reorganize in a way which ensures conservative
  transport of material down the pile.  If the height of the pile is
  represented by a quantity $h(x)$, then this model may be summarized
  by an equation of motion of the form
\begin{displaymath}
\dot h = f(h,\nabla h) + \nabla\xi.
\end{displaymath}
Here, $f(h,\nabla h)$ is some function which enforces the conservative
transport and $\xi$ represents the response of the system to the coherent
noise (which varies with event size $s$ approximately as $s^{-2}$).}

A second variation of interest is what we have termed the ``multi-trait''
version of the model, after a similar generalization of the Bak-Sneppen
model proposed by Boettcher and Paczuski~\cite{BP96}.  There are some
systems, such as stock markets or biological extinctions, which one might
be tempted to model using a dynamics of the type we have been discussing,
except that these systems are clearly subject to not just one, but many
different kinds of external stress.  The multi-trait version of the model
takes this into account by introducing a number $M$ of different stress
types, denoted $\eta_k$, or alternatively an $M$-dimensional vector stress.
The threshold also becomes a vector ${\bf x}$, and the rules of the
dynamics are modified so that an agent $i$ moves if any one of the
components of ${\bf x}_i$ is less than the corresponding component of the
stress vector.  To a first approximation, one can treat the probability
$p_{\rm move}({\bf x})$ of an agent moving in such a system as the
probability that the stress level exceeds the lowest component of the
threshold vector.  In this case, the multi-trait version of the model is
identical to the single-trait version but with a different choice for
$p_{\rm thresh}(x)$ (one which reflects the probability distribution of the
lowest of $M$ random numbers).  A slightly more sophisticated treatment
equates the probability that an agent will move to the probability that any
one of the components $x_k$ of its threshold vector will be exceeded by the
corresponding stress level:
\begin{equation}
1 - p_{\rm move}({\bf x}) = \prod_k [1 - p_{\rm move}(x_k)]
                           \approx 1 - \sum_k p_{\rm move}(x_k).
\end{equation}
Either way the result is the same: we expect robust power-law distributions
in the multi-trait model just as we had in the single-trait version, and
numerical experiments bear this out.

\section{Conclusions}
\label{conclude}
We have introduced a new class of model in which a large number of agents
organize under the action of an externally imposed coherent stress.
Simulations indicate that these models generate robust distributions of
sizes of reorganization events, agent lifetimes and aftershock events,
regardless of the values of a variety of microscopic parameters, within
certain limits.  We have given an approximate solution of these models
which explains how this robustness arises, and what its limits are.  All
indications are that for virtually any distribution of driving noise which
one might expect to encounter in nature, coherent-noise driven systems of
this type may be expected to display power-law response distributions.  We
believe that this mechanism may be an important step towards the realistic
modelling of intermittency and collective behaviour in large systems, and
as such it may open the way for new insights into characteristic behaviours
which seemingly are at odds with the central limit theorem.

\section{Acknowledgements}
One of us (KS) would like to thank the Santa Fe Institute for their
hospitality and financial support while this work was carried out.  This
research was funded in part by the Santa Fe Institute and DARPA under grant
number ONR N00014--95--1--0975.

\end{document}